\documentstyle[12pt]{article}

\def\be{\begin{equation}}
\def\ee{\end{equation}}
\def\ben{\begin{displaymath}}
\def\een{\end{displaymath}}
\def\ba{\begin{array}{c}}
\def\bal{\begin{array}{l}}
\def\ea{\end{array}}
\def\p{\partial}
\begin{document}

 \begin{center}
.

\vspace{.35cm}

{\Large \bf
Maximal couplings in ${\cal PT}-$symmetric chain-models with the
real spectrum of energies}
\end{center}

\vspace{10mm}

 \begin{center}

 {\bf Miloslav Znojil}

 \vspace{3mm}
Nuclear Physics Institute ASCR,

 250 68 \v{R}e\v{z}, Czech Republic

{e-mail: znojil@ujf.cas.cz}

\vspace{3mm}

\vspace{5mm}


\end{center}

\vspace{5mm}

\section*{Abstract}
%
%
%
%

The domain ${\cal D}$ of all the coupling strengths compatible
with the reality of the energies is studied for a family of
non-Hermitian $N$ by $N$ matrix Hamiltonians $H^{(N)}$ with
tridiagonal and ${\cal PT}-$symmetric structure. At all dimensions
$N$, the coordinates are found of the extremal points at which the
boundary hypersurface $\partial {\cal D}$ touches the
circumscribed sphere (for odd $N=2M+1$) or ellipsoid (for even
$N=2K$).

\section{Introduction
}

\subsection{Non-Hermitian chain models}

In many quantum systems (typically, in nuclear and condensed
matter physics), the observed spectra can be fitted by the
equidistant harmonic-oscillator energies (i.e., by  $E^{(HO)}_n
=2n+1$ in suitable units). An improvement of this fit can be based
on a perturbatively mediated transition, say, to the popular
nearest-neighbor-interaction model with an infinite-dimensional
``chain-model" tridiagonal Hamiltonian
 \be
 H^{(\infty)}=\left [\begin {array}{ccccc} 1&a_0&0&\ldots&\\
 \noalign{\medskip}b_0&
 3&a_1&0&\ldots\\
 \noalign{\medskip}0&b_1&5&a_2&\ddots
 \\
 \noalign{\medskip}\vdots&\ddots&\ddots&\ddots&\ddots
 \end {array}\right ]\,.
 \label{NNI}
 \ee
For the real coupling strengths $a_n$, $b_m$ the manifest
asymmetry of our Hamiltonian $ H^{(N)}$ (with infinite as well as
finite matrix dimension $N$) need not necessarily contradict the
postulates of Quantum Mechanics. This may be illustrated on the so
called Swanson's model with $N=\infty$ \cite{Swanson} or, more
easily, on the simplest truncated two-dimensional special case of
eq.~(\ref{NNI}),
 \ben
 H^{(2)}=\left (
 \begin{array}{cc}
 1&a\\
 b&3
 \ea
 \right )\,.
 \een
{\em Both} its eigenvalues $E_\pm = 2\pm \sqrt{1+ab}$ remain real
(i.e., in principle, ``observable") whenever $ab\geq -1$. Inside
this ``domain of physical acceptability", i.e., for
 \be
 (a,b) \in {\cal D}^{(2)}\ \equiv\
 \left \{
 (x,y)\, \right |
 \, x,y \in I\!\!R, \ xy>-1
 \left .
 \right \}
 \ee
these energies are also non-degenerate. This gives the technically
welcome guarantee that $H^{(2)}$ can be diagonalized in a
bi-orthogonal basis formed by the two respective sets of the right
and left eigenvectors $|\pm \rangle$ and $|\pm \rangle\! \rangle$
such that
 \be
 H^{(2)}\,|\pm \rangle =E_\pm \, |\pm \rangle, \ \ \ \ \ \
 \langle\! \langle \pm |\,
 H^{(2)}=
 \langle\! \langle \pm |\ E_\pm\,.
 \label{gerd}
 \ee
The diagonalizability is lost on the boundary $\p {\cal D}^{(2)}$
[where the basis of eq.~(\ref{gerd}) becomes incomplete] and the
reality of the energies is lost everywhere in the open complement
of ${\cal D}^{(2)}$.

In a way discussed thoroughly in our recent letter \cite{Hendrik}
the two-dimensional model (\ref{gerd}) proves particularly useful
for an elementary explicit illustration of one of the ``key
tricks" which re-assigns the necessary Hermiticity to the similar
operators. The goal is being achieved by a suitable redefinition
of the metric $\Theta$ and, hence, of the scalar product,
  \be
 |\,\psi\rangle \odot |\,\phi \rangle \ \equiv\ \langle \psi|\,
 \Theta\,|\,\phi \rangle
 \,,\ \ \ \ \ \ \Theta=\Theta^\dagger>0\,.
 \label{nevwpr}
 \ee
During the last few years, such a recipe has been revealed and/or
implemented by several independent groups of authors sampled here
in refs.~\cite{Geyer,BBJ,ali,ja}.

\subsection{Construction of the metrics $\Theta$ for a given
Hamiltonian}

One should re-emphasize that in general, a redefinition of the
metric $\Theta$ in Hilbert space is fully compatible with the
postulates of Quantum Mechanics, provided only that the reality of
the spectrum is guaranteed. From time to time, the efficiency of
the trick is being confirmed in various less standard applications
of Quantum Theory~\cite{Reed,Reedb}.

In the notation of eq.~(\ref{gerd}) the essence of the trick
derives from the observation that for many manifestly
non-Hermitian Hamiltonians $H \neq H^\dagger$ with real spectra
one can follow the two-dimensional guidance and construct the two
families of the left eigenvectors $|n \rangle\! \rangle$ and of
the right eigenvectors $|m \rangle$ of a given $H$. They form a
biorthogonal basis in Hilbert space. In the next step one easily
verifies, in all the finite-dimensional cases at least, that the
operator defined by the spectral-representation-like formal
expansion
 \be
 \Theta=\sum_n\,|n
 \rangle\! \rangle\,s_n\,\langle\! \langle n |
 \label{newpr}
 \ee
satisfies the linear operator equation
 \be
 H^\dagger\,\Theta = \Theta\,H\,.
 \label{metrica}
 \ee
In the final step of the argument one restricts {\em all} the
parameters $s_n$ to the real and positive numbers and concludes
that the properties of the resulting operator $\Theta$ qualify it
for a metric-operator interpretation as discussed in the review
paper \cite{Geyer}. This means that the ``correct" inner product
is ambiguous as it may be defined by {\em any} prescription
(\ref{newpr}). Its choice in fact fixes our selection of an
explicit representation of the Hilbert space of states and, hence,
``the physics".

There exist several remarkable differences between the unique,
``standard" choice of $\Theta=I$ and all the ``nonstandard"
$\Theta \neq I$ in eq.~(\ref{nevwpr}). For this reason, usually,
the Hermiticity condition (\ref{metrica}) with $\Theta \neq I$ is
being re-named to ``quasi-Hermiticity" \cite{Geyer,ali}. One of
the most characteristic consequences of the quasi-Hermiticity of a
Hamiltonian $H$ lies in the necessity of a specification of the
domain ${\cal D}$ of parameters where the spectrum of energies
remains real.

\section{${\cal PT}-$symmetric models }

\subsection{Modified harmonic oscillators}

Under the assumption $H \neq H^\dagger$ some of the eigenvalues
become complex whenever we leave the quasi-Hermiticity domain
${\cal D}$ of parameters in $H$. In the context of one-dimensional
differential Schr\"{o}dinger operators the problem has been made
popular by Bender et al \cite{BBJ} who studied the generalized
Bessis' oscillators
 \be
  H^{(GB)}(\nu)=-\frac{d^2}{dx^2}+g(x)\,x^2\,,
  \ \ \ \ \ \ \ \ g(x)=({\rm i}x)^\nu\,,
  \ \ \ \ \ \ \ \ \  \nu \in I\!\!R
  \label{Bessis}
  \ee
and conjectured that all the bound-state energies remain real iff
$\nu \geq 0$, i.e., inside the very large domain ${\cal D}^{(GB)}
\equiv (0,\infty)$ of the exponents $\nu$. Rigorously, this
conjecture has only been proved three years later \cite{DDT}. One
should notice that the difficulty of this proof is in a sharp
contrast with the elementary character of the above-mentioned
construction of ${\cal D}^{(2)}$ related to the finite-dimensional
$H^{(2)}$.

Our present paper is inspired by the question of feasibility of
the constructions of the quasi-Hermiticity domains ${\cal
D}^{(N)}$ for matrices at the higher dimensions $N>2$.
Predecessors of such a project can be seen not only in the
exhaustive analyses of virtually all the two-dimensional cases
\cite{turek} but also in our recent note \cite{pierot} where we
reported the feasibility of a complete and non-numerical
reconstruction of the domain ${\cal D}^{(3)}$ for certain special
${\cal PT}-$symmetric three by three toy Hamiltonians.

We shall address here the natural question of the specification of
${\cal D}$ for the matrix family of the perturbed harmonic
oscillator Hamiltonians (\ref{NNI}) restricted by an additional
requirement of their ${\cal PT}-$symmetry. We believe that such
band-matrix models are really exceptional. One of our reasons
originates from the observation that in the most elementary
differential-operator representation of $H^{(HO)}$, all the wave
functions $\psi_n(x)$ pertaining to the above-listed energies
$E_n^{(HO)}=2n+1$ are endowed with an additional, parity quantum
number, ${\cal P}\,\psi_n(x) = \psi_n(-x) = (-1)^n\,\psi_n(x)$.
This is a consequence of the commutativity ${\cal P}\, H^{(HO)}
=H^{(HO)} \,{\cal P}$ which is manifestly broken in all the
perturbed matrix models $H^{(N)}$.

The ${\cal PT}-$symmetry requirement ${\cal
PT}\,H^{(N)}=H^{(N)}\,{\cal PT}$ is quite natural to impose,
especially because the operator ${\cal T}$ can be treated as a
mere transposition. In addition, it is easy to imagine that in the
given basis the operator ${\cal P}$ is represented by the diagonal
matrix with elements ${\cal P}_{nn}=(-1)^n$. As a net consequence,
the requirement of the ${\cal PT}-$symmetry degenerates to the
elementary rule $a_n=-b_n$ at all subscripts $n$ in
eq.~(\ref{NNI}),
 \be
 H^{(N)}=\left [\begin {array}{ccccc} 1&a_0&0&\ldots&0\\
 \noalign{\medskip}-a_0&
 3&a_1&\ddots&\vdots\\
 \noalign{\medskip}0&-a_1&5&\ddots&0
 \\
 \noalign{\medskip}\vdots&\ddots&\ddots&\ddots&a_{N-2}\\
 0&\ldots&0&-a_{N-2}&2N-1
 \end {array}\right ]\,.
 \label{NNIPT}
 \ee
These are the models which we are going to analyze.

\subsection{An additional ``up-down" symmetrization}

\subsubsection{A generic
 attraction of the levels. \label{uola}}

After a few numerical experiments with eq.~(\ref{NNIPT}) one
reveals a comparatively robust survival of the reality of the
spectrum in perturbative regime. The phenomenon can be understood
as one of the mathematically most interesting consequences of the
``sufficient separation" of the matrix elements $1,3,\ldots$ along
the main diagonal \cite{Graffi}.

In contrast, even a strict observation of the equidistance of the
elements on the main diagonal need not necessarily be of any help
in a deeply non-perturbative regime. This danger is well known and
the monograph \cite{Wilkinson} can be consulted for an extremely
persuasive illustration of the emergence of unexpected
difficulties even in a {\em symmetric} nonperturbative version of
our example~(\ref{NNIPT}) with an apparently innocent choice of
the dimension $N=20$ and with an apparently ``not too
nonperturbative" constant diagonal where
$a_0=a_1=\ldots=a_{18}=a$.

In order to avoid similar complications in our present ${\cal
PT}-$symmetric models we may try to assume, in the first step,
that just a single coupling $a=a_k$ becomes large. In such a case,
solely the two neighboring energies become involved and
perceivably modified. Generically, in a way controlled by the mere
two-dimensional submatrix $H^{(2)}_{(k)}$ of $H^{(N)}$ we have
 \ben
 \ \ \ \ \ \ \ \ \ \ \ \ \ \ \ \
 H^{(2)}_{(k)}=\left (
 \begin{array}{cc}
 2k+1&a\\
 -a&2k+3
 \ea
 \right )\,
 \een
so that the energy values become ``attracted" by each other in
proportion to $|a_k|$  at any $k<N-1$,
 \ben
 E_k = 2k+2- \sqrt{1-a^2}\,,\ \ \ \ \ \ \
 E_{k+1} = 2k+2+ \sqrt{1-a^2}\,.
 \een
The mechanism of this effect is virtually independent of the rest
of the spectrum (which may be considered pre-diagonalized) so that
we may {\em always} expect that some energies get complexified
{\em whenever} the couplings become sufficiently strong.

This means that, intuitively, we may always visualize the
coupling-dependence of the energies as their {\em mutual}
attraction. In this sense we are able to guess that the levels
$E_{n_0}$ in the middle of the matrix (i.e., such that $n \approx
n_0 \approx N/2$) will be ``maximally protected" against the
complexification due to their multiple and balanced ``up" and
``down" attraction by all the other levels.

Such a balance may be quantitatively (though not qualitatively)
violated by the differences in the absolute values of the pairs of
couplings $a_{n_0+k}$ and $a_{n_0-k}$ at all the allowed
index-shifts $k$. For this reason we shall restrict our present
attention to the special  class of the Hamiltonian
matrices~(\ref{NNIPT}) which are, in this sense, symmetrized and
have
 \be
 a_{j}=a_{N-2-j}\, \ \ \ \ \ \ j = 0, 1, \ldots, j_{max}\ (\
 = {\it entier} [N/2]\,)\,.
 \label{syme}
 \ee
This means that everywhere in what follows we shall reduce the
class of the $(N-1)-$parametric chain models~(\ref{NNIPT}) to its
``up-down-symmetrized" subset
 \be
 H^{(N)}
 =\left [\begin {array}{ccccc}
  \delta-1&a_0&0&\ldots&0\\
 -a_0& \delta-3&\ddots&\ddots&\vdots\\
 0&-a_1&\delta-5&a_1&0
 \\
 \vdots&\ddots&\ddots&\ddots&a_{0}\\
 0&\ldots&0&-a_{0}&\delta-2N+1
 \end {array}\right ]\,.
 \label{NNIPTS}
 \ee
Obviously, the specific choice of the global shift $\delta=N$ of
the origin of the energy scale makes also the main diagonal of the
whole matrix ``up-down" symmetric. Still, it is slightly
unpleasant that at the strictly integer half-dimensions $K=N/2$,
the ``last" free parameter $a_{j_{max}}$ enters our matrix $
H^{(N)} $ ``anomalously", i.e., just twice. This means that the
parity of $N$ introduces a fairly nontrivial difference between
the corresponding up-down-symmetric models~(\ref{NNIPTS}).

\subsubsection{Even dimensions $N=2K$, $K=1, 2, \ldots$. }

As long as we intend to analyze the secular determinants of our
matrices \cite{Wilkinson}, it makes sense to simplify our notation
and, in particular, to get rid of the subscripts and abbreviate
$a=a_{j_{max}}$, $b = a_{j_{max}-1}$ and so on, up to the last
element $a_0$ abbreviated, whenever needed, by the last letter
$z$. In this notation, obviously, the symbol $z=a_0$ coincides
with $a=a_{j_{max}}=a_0$ at $K=1$ in
 \ben
 H^{(2)} = \left [\begin {array}{cc} 1&a\\{}-a&-1\end {array}\right ]
 \een
or with $b$ at $K=2$ in
 \ben
 H^{(4)} = \left [\begin {array}{cccc}
  3&b   &0  &0\\
 -b&1   &a  &0\\
  0&-a  &-1 &b\\
  0&0&-b&-3
 \end {array}\right ]
 \een
etc. We see that the general matrix~(\ref{NNIPTS}) with $N=2K$ may
be easily understood as partitioned into four $K-$dimensional
submatrices,
 \ben
 H^{(2K)}=\left [\begin {array}{cccc|cccc}
  2K-1&z&0&\ldots&&&&
  \\
  -z&\ddots&\ddots&\ddots&\vdots&&&
 \\
  0&\ddots&3&b&0&\ldots&&
 \\
 \vdots&\ddots&-b&1&a&0&\ldots&
 \\
 \hline
 &\ldots&0&-a&-1&b&0&\ldots
  \\
 &&\ldots&0&-b &-3&\ddots&
 \\
 &&&\vdots&\ddots&\ddots &\ddots&z
 \\
 &&&&\ldots&0&-z&1-2K\end
 {array} \right ]
 \een
The simplest illustrative example $H^{(2)}$ has already been
shortly discussed above (cf. also \cite{Hendrik}). In the general
case the secular polynomial $\det (H^{(2K)}-E)$ will be a
polynomial of the $K-$th degree in $s=E^2$ and it will only depend
on the squares of the couplings $a_{j_{max}}^2\equiv a^2=A$,
$a_{j_{max}-1}^2\equiv b^2=B$, $\ldots$, $a_0^2 \equiv z^2 =Z$.

\subsubsection{Odd dimensions $N=2M+1$, $M=1,2, \ldots$. \label{lisi} }

Whenever the dimension of our band-matrix Hamiltonian $H^{(N)}$
with equidistant matrix elements on its main diagonal is odd,
$N=2M+1$, we have
 \ben
 H^{(2M+1)}=\left [\begin {array}{ccc|c|ccc} 2M&z&0&0&0&0&0\\
 {}-z&
 \ddots&\ddots&0&0&0&0\\
 {}0&\ddots&2&a&0&0&0
 \\
 \hline
 {}0&0&-a&0&a&0&0
 \\
 \hline
 {}0&0&0&-a&-2&\ddots&0\\
 {}0&0&0&0&\ddots&\ddots&z
 \\
 {}0&0&0&0&0&-z&-
 2M\end {array}\right ]\,.
 \een
Here the central matrix element vanishes and the ${\cal
PT}-$symmetric coupling is mediated again by the $M$ real matrix
elements $a, b, \ldots, z$. Omitting the overall factor $E$ we may
reduce the secular polynomial $\det (H^{(2M+1)}-E)$ to a
polynomial of the $M-$th degree in $s=E^2$. It will again depend
on the squares of the couplings only.

\section{Hamiltonians of the even dimensions $N=2K$ }

In the two-dimensional case with $K=1$ the whole discussion
remains entirely elementary (see above) and one can conclude that
there exist precisely two points of the boundary $\p {\cal
D}^{(2)}$ (called ``exceptional points" in the literature
\cite{Kato,Heiss}) which are defined by the elementary rule
$a^{(EP)}_{\pm} =\pm 1$, i.e., by the single root $A^{(EP)}=1$ of
the single energy-degeneracy condition.

\subsection{Four by four model, $K=2$ \label{ctyrak} }

For the four by four Hamiltonian $H^{(4)}$ the standard definition
of the spectrum
 \ben
 \det \left [\begin {array}{cccc} 3-{\it  E}&b&0&0\\{}-b&1-{
\it  E}&a&0\\{}0&-a&-1-{\it  E}&b\\{}0 &0&-b&-3-{\it  E}\end
{array}\right ]=0
 \een
i.e., the quadratic secular equation for $s=E^2$,
 \ben
 {s}^{2}+\left (-10+2\,{b}^{2}+{a}^{2}\right )s+9+6\,{b}^{2}-9\,{a}^{2}
 +{b}^{4}=0
 \een
can easily be solved in closed form,
 \be 
 s=s_\pm =5-{b}^{2}-1/2\,{a}^{2}\pm 1/2\,\sqrt {64-64\,{b}^{2}+16\,{a}^{2}+4\,{b}^{
 2}{a}^{2}+{a}^{4}}\,.
 \label{rroty}
 \ee
These formulae may be read as an implicit definition of ${\cal
D}^{(4)}$, i.e., of the reality domain of the energies or,
equivalently \cite{ali}, of the quasi-Hermiticity domain of the
Hamiltonian of our $K=2$ chain model.

For a more explicit construction of ${\cal D}^{(4)}$ we can make
use of the up-down symmetry (\ref{syme}) and imagine that during
the initial perturbative mutual attraction of the neighboring
levels one can only guarantee the growth of the ground-state
minimum $E_{0}=E_{-,+}\ \equiv\ -\sqrt{s_+}$ and the decrease of
the top-state maximum $E_{3}=E_{+,+}\ \equiv\ +\sqrt{s_+}$.

Beyond perturbative domain, at certain ``exceptional-point"
combinations $(a,b)=(a,b)^{(EP_+)}$ of the sufficiently large
strengths $a$ and $b$, the latter two extreme energy levels  will
ultimately coincide (and, immediately afterwards, complexify) in a
way discussed in paragraph \ref{uola} above,
$E_{-,+}^{(EP)}=E_{+,+}^{(EP)}=0$. At another set of the EP
coupling doublets $(a,b)=(a,b)^{(EP_-)}$, both the two ``internal"
levels  may also coincide as well, $E_{\pm,-}\ \equiv\ \pm
\sqrt{s_-}=0$. In this way, the complete boundary $\p {\cal
D}^{(4)}$ of the quasi-Hermiticity domain is a curve in the $a-b$
plane formed by the ``weaker" doublets of the EP-strenghs
$(a,b)^{(EP_\pm)}$. The shape of such a boundary can be deduced
from eq.~(\ref{rroty}) (cf. Figure~1).

In a magnified detail, Figure~2 demonstrates that the graphical
representation ceases to be reliable in the fairly large vicinity
of the common maximum of the sizes of the allowed couplings $a$
and $b$. Fortunately, near any such a ``extremely exceptional"
point $(a,b) = \{ \, (\pm \sqrt{A^{(EEP)}},\pm \sqrt{B^{(EEP)}}
\}$ of the $a-b$ plane,  the details of the shape of the boundary
$\p {\cal D}^{(4)}$ can be described by the purely analytic means.
An extension of the latter observation to all the dimensions $N$
will become, after all, a core of our present message.

Let us explain the method for $N=2K$ at any $K$. In the first step
one realizes that $s^{(EEP)}=0$ due to the up-down symmetry. As
long as this must be the only root (i.e., a maximally degenerate
root) of the polynomial secular equation
 \be
 s^{K}+ P_{K-1}(A,B,\ldots)\,s^{K-1}+
 P_{K-2}(A,B,\ldots)\,s^{K-2}+\ldots=0
 \label{sekurita}
 \ee
it can only exist if the $K$ values $A^{(EEP)},B^{(EEP)},\ldots$
of the EEP coupling strengths satisfy the nonlinear set of the
following $K$ necessary conditions,
 \ben
 P_{K-1}\left (A^{(EEP)},B^{(EEP)},\ldots\right )=0, \ \ \ \ \ \
 \een
 \ben
 P_{K-2}\left (A^{(EEP)},B^{(DEEP)},\ldots\right )=0,
 \een
 \be
 \ \ \ \ldots
 \label{groebb}
 \ee
 \ben
 P_0\left (A^{(EEP)},B^{(EEP)},\ldots\right )=0\,.
 \een
At $K=2$ the latter set of polynomial equations reads
 \ben
 A+2\,B=10, \ \ \ \ \ \ \ \ \
 (3+B)^2
 =9\,A
 \een
and an elimination of $A$ leads to a quadratic equation for $B+3$
giving a spurious solution $A=64$ and $B=-27$ (which would imply
an imaginary coupling $b$) and the unique correct solution
$A^{(EEP)}=4$ and $B^{(EEP)}=3$.

\subsection{Six by six model, $K=3$.}

In a full parallel with the preceding subsection, secular equation
 \ben
 \det \left [\begin {array}{cccccc} 5-{\it
  E}&c&0&0&0&0\\{} -c&3-{\it  E}&b&0&0&0\\{}0&-b&1-{\it  E}&a&0&0
\\{}0&0&-a&-1-{\it  E}&b&0\\{}0&0&0&-b
&-3-{\it  E}&c\\{}0&0&0&0&-c&-5-{\it  E}\end {array} \right ]=0
 \een
in its polynomial form (\ref{sekurita}),
%
%
 \ben
 {s}^{3}+\left (2\,{b}^{2}-35+2\,{c}^{2}+{a}^{2}\right ){s}^{2}+
  \een
  \ben
  +\left (
{b}^{4}+2\,{c}^{2}{a}^{2}-44\,{b}^{2}+28\,{c}^{2}-34\,{a}^{2}+{c}^{4}+
259+2\,{b}^{2}{c}^{2}\right
)s+
 \een
 \ben
 +{a}^{2}{c}^{4}-10\,{b}^{2}{c}^{2}+30\,{
c}^{2}{a}^{2}+225\,{a}^{2}-30\,{c}^{2}-{c}^{4}-25\,{b}^{4}-225-150\,{b
}^{2}=0
 \een
remains solvable in closed form. As long as our present attention
is concentrated on the EEP extremes, we shall skip the details of
the complete description of the hedgehog-shaped surface $\p {\cal
D}^{(6)}$ in the full three-parametric space and note only that
this shape must be all contained within the ellipsoid with the
boundary described by the first constraint of eq.~(\ref{groebb}),
${a}^{2}+2\,{b}^{2}+2\,{c}^{2}=35$.

At $K=3$ the full solution of the triplet of eqs.~(\ref{groebb})
ceases to be easy but it still remains feasible. Besides the
unique and acceptable correct solution
 \be
 A^{(EEP)}=9\,,\ \ \ B^{(EEP)}=8\,,\ \ \ C^{(EEP)}=5\,,\ \ \ \ \ \
 \ \ \ K =3,
 \label{sol3}
 \ee
one obtains another set of the alternative solutions generated,
after the patient elimination of $A$ and $B$, in terms of roots of
a final ``effective" polynomial in single variable~$C$,
 \be
 416\,{C}^{4}+20909\,{C}^{3}+22505\,{C}^{2}+
28734375\,{C}^{}-48828125=0. \label{sekul3} \ \ \ \ \ \ \ \ K=3\,.
 \ee
Out of its two real roots, $C_- = -65.80360706$ and $C_+
=1.693394621$, the former one is manifestly spurious giving the
imaginary coupling $c$. For the latter root we have to recall the
corresponding condition
 \ben
 22156250\,B_+ +2912\,{C_+ }^{3}+1446363\,{C_+ }^{2}+820546875+9654410\,
{C_+ }^{}=0
 \een
to see that the coupling $b=\sqrt{B_+}$ is imaginary and should be
rejected as spurious as well.

%
%

\subsection{Eight by eight model, $K=4$.}

Out of the four EEP constraints (\ref{groebb}) at $K=4$ the first
equation $P_3(a^2,b^2,c^2,d^2)=0$ defines the surface of an
ellipsoid or, after the change of variables $a\to A=a^2$ etc, a
planar side of a simplex,
 \ben
 A+2\,B+2\,C+2\,D=84\,.
 \een
By construction, the domain ${\cal D}^{(8)}$ is circumscribed by
this ellipsoid or simplex. Unfortunately, one hardly finds any
immediate geometric interpretation of the remaining quadratic,
cubic and quartic polynomial equations $P_2(A,B,C,D)=0$,
$P_1(A,B,C,D)=0$ and $P_0(A,B,C,D)=0$ containing 13, 19 and 20
individual terms, respectively, and admitting just marginal
simplifications, e.g., to the 9-term equation
 \ben
 1974+(B+C+D)^2+
 2\,AD+2\,BD+2\,AC
 =83\,A+142\,B+70\,C-50\,D\,
 \een
in the $P_2-$case, etc.

In this setting it comes as a real surprise that the above-derived
$K=3$ rule (\ref{sol3}) still finds its unique $K=4$ counterpart
which, in addition, possesses the closed form again,
 \be
 A^{(EEP)}=16\,,\ \ \ B^{(EEP)}=15\,,\ \ \ C^{(EEP)}=12\,,\ \ \
 D^{(EEP)}=7\,, \ \ \ \ \ \ \ K =4\,.
 \label{sol4}
 \ee
Its derivation necessitated the use of the fully computer-assisted
Groebner-basis technique. Just for illustration one may mention
the $K=4$ form of the final ``effective" polynomial equation,
 \ben
 314432\,D^{17}-5932158016\,D^{16}+ \ldots 
+153712881941946532798614648361265167=0,
 \een
representing the ``next-door neighbor" of the still exactly
factorizable eq.~(\ref{sekul3}).

In a test of the uniqueness of solution (\ref{sol4}) one finds out
that it possesses seven real and positive roots $D$. Out of them,
the following three ones are negative and, hence, manifestly
spurious,
 $
-203.9147095, -156.6667001, -55.49992441
 $.
We skipped the proof of the spuriosity for the remaining four
roots, viz., of $0.4192854385, 5.354156128, 1354.675195$ and
$18028.16789$ since the related calculations, however
straightforward, become unpleasant and clumsy. For example, the
values of $A$ are given by the rule $\alpha\,\times \,A =$ (a
polynomial in $D$ of 16th degree) where the number of digits in
the auxiliary integer constant $\alpha$ exceeds one hundred.

\subsection{Arbitrary even dimension $N=2K$.}

Even though we did not dare to test the applicability of the
Gr\"{o}bner-basis technique at $K=5$, we were lucky in noticing
that the previous results already admitted the following
extrapolation {\em to any $K$},
 \be
 A^{(EEP)}=K^2,\   B^{(EEP)}=K^2-1^2,\   C^{(EEP)}=K^2-2^2,\
 D^{(EEP)}=K^2-3^2, \  \ldots\,.
 \label{solka}
 \ee
This is our first main result. The validity of this empirically
revealed rule has subsequently been tested and confirmed by the
incomparably simpler direct insertions.

As a byproduct of these verifications, the general elipsoidal
surface form of the first item in eq.~(\ref{groebb}) has been
predicted from the data available at $K\leq 4$ and re-confirmed at
several higher $K>4$ giving, in terms of the original
coupling-strength variables of eq.~(\ref{NNIPT}) with symmetry
(\ref{syme}),
 \be
 A+2\,
 \left (B+C+\ldots +Z
 \right)
 \ \equiv\
 a^2_{j_{max}}+2\,a^2_{j_{max}-1}+\ldots +2\,a_0^2=
 \sum_{k=0}^{N-2}\,a^2_k
 =\frac{4\,K^3-K}{3}\,
 \label{star}
 \ee
or, in the form of an immersion of ${\cal D}$ in an ellipsoid in
$K$ dimensions,
 \be
 a^2+2\,b^2+\ldots +2\,z^2 \ \equiv \
 \sum_{k=0}^{N-2}\,a^2_k
 \leq
 \frac{4\,K^3-K}{3}\,.
  \label{stardustbinec}
 \ee
These observations are in a complete agreement with the
individually evaluated formulae and carry a geometric
interpretation showing that every domain ${\cal D}^{(2K)}$ (where
all the energies remain real) is circumscribed by a certain
elipsoidal hypersurface. Its intersections with the boundary $\p
{\cal D}^{(2K)}$  coincide with the $2^K$ EEP points with the
coordinates $ a^{(EEP)} =\pm K$, $\ b^{(EEP)}=\pm \sqrt{K^2-1}$
etc.

\section{Hamiltonians of the odd dimensions $N=2M+1$}

In a one-parameteric three-by-three illustration with $M=1$,
 \ben
 H^{(3)} = \left [\begin {array}{ccc} 2&a&0\\{}-a&0
 &a\\{}0&-a&-2\end {array}\right ]
 \een
the determination of the interval of quasi-Hermiticity $a \in
{\cal D}^{(3)}=(-\sqrt{2},\sqrt{2})$ is trivial since the secular
equation $-E^{3}+\left (4-2\,{a}^{2}\right )E=0 $ is exactly
solvable. In a remark~\cite{pierot} we also studied a ``generic"
three-dimensional (and three-parametric) matrix model where we
relaxed both the equidistance assumption concerning the main
diagonal {\em and} our present simplifying ``up-down"
symmetrization assumption $a_0=a_1$.

\subsection{Five by five model, $M=2$.}

A comparatively elementary two-parametric example of our present
class of models of section \ref{lisi} is still encountered at
$M=2$,
 \ben
 H^{(5)}=\left [\begin {array}{ccccc} 4&b&0&0&0\\{}-b&
2&a&0&0\\{}0&-a&0&a&0
\\{}0&0&-a&-2&b\\{}0&0&0&-b&-
4\end {array}\right ]\,.
 \een
Its secular equation gives the central constant energy $E_0=0$.
The other two pairs of the real or complex conjugate levels
$E_{n}=-E_{-n}=\sqrt{s}$ with $n=1,2$ are obtained from the
remaining polynomial equation in the new variable $s=E^2$,
  \be
  -s^2+\left (20-2\,{b}^{2}-2\,{a}^{2}\right )s-
 64-16\,{b}^{2}+32\,{a}^{2}-{b}^{4}-2\,{a}^{2}{b}^{2}=0\,.
 \label{tojeon}
 \ee
We should determine the domain ${\cal D}^{(5)}$ in which all the
energies remain real. This means that inside the closure of the
domain of quasi-Hermiticity ${\cal D}^{(5)}$ both the roots of
eq.~(\ref{tojeon}) must be non-negative.

Our task is elementary since the $M=2$ eigenvalue problem is
solvable in closed and compact form,
 \ben
 E_{\pm 1}=\pm \sqrt{10-{a}^{2}-{b}^{2}-\sqrt
 {36+12\,{a}^{2}+{a}^{4}-36\,{b}^{2}}}\,,
 \een
 \ben
 E_{\pm 2}=\pm \sqrt{10-{a}^{2}-{b}^{2}+\sqrt
 {36+12\,{a}^{2}+{a}^{4}-36\,{b}^{2}}}\,.
 \een
Thus, the results of the method of preceding section may be
complemented by direct calculations. In terms of the two
non-negative quantities $A=a^2\geq 0$ and $B=b^2\geq 0$ the
reality of the energies will be guaranteed by the triplet of
inequalitites. The first one reads $10\geq A+B$ and restricts the
allowed values of $A$ and $B$ to a simplex. The second condition
$36+12\,A+{A}^{2} \geq 36\,B$ requires that the allowed values of
$B$ must lie below a growing branch of a parabola
$B_{max}=B_{max}(A)$. The third condition $ (8+B)^2 \geq
(32-2\,B)\,A$ represents an easily visualized upper bound for
$A\leq A_{max}=A_{max}(B)$ where the latter hyperbola-shaped
function grows with $B$ in all the interval of interest.

Beyond the above direct proof we may also parallel the
considerations of the preceding section and imagine that the
symmetry of eq.~(\ref{tojeon}) implies that its triple root must
vanish, $s=s^{(EEP)}=0$. This means that in the polynomial
eq.~(\ref{tojeon}) both the coefficients at the subdominant powers
of $s$ must vanish. These two coupled conditions degenerate to the
single quadratic equation with the unique non-spurious solution
$A^{(EEP)}=6$ and $B^{(EEP)}=4$. Thus, in a way complementing our
above direct discussion of the reality of the energies we see that
at our EEP point all the three above-mentioned inequalities become
saturated simultaneously.

\subsection{Seven by seven model, $M=3$.}

By the same Gr\"{o}bner-basis method as above we derive the result
 \be
 A^{(EEP)}=12\,,\ \ \ B^{(EEP)}=10\,,\ \ \ C^{(EEP)}=6\,,\ \ \
 \ \ \ M=3\,.
 \label{so3}
 \ee
It is again unique because one of the two roots $C_\pm=27 \pm
9\,\sqrt {21}$ of the ``first alternative" Gr\"{o}bnerian
``effective" equation ${{\it C}}^{2}-54\,{\it C}=972$ and {\em
both} the roots $-354 \pm 60\,\sqrt {34} $ of the ``second
alternative" equation ${{\it C}}^{2}+708\,{\it C}+2916=0$ are
negative while the only remaining positive root $C_+=68.24318125$
gives the negative ${\it {B}}=28-3 \,{\it C}$.

\subsection{All the $(2M+1)-$dimensional models with $M\geq 4$.}

At $M=4$ we still were able to evaluate the explicit form of the
secular equation,
 \ben
 14745600-7372800\,{\it {A}}+ \ldots +
\left (-2\,{\it C}+220-2\,{\it {B}}-2\,{\it {A}}-2\,{\it ff
}-2\,{\it D}\right ){s}^{4}-{s}^{5}=0
 \een
and we also still {\em computed} the $M=4$ EEP solution directly,
 \be
 A^{(EEP)}=20\,,\ \ \ B^{(EEP)}=18\,,\ \ \ C^{(EEP)}=14\,,\ \ \
 D^{(EEP)}=8\,, \ \ \ \ \ \ \ M =4\,.
 \label{s4}
 \ee
We already gave up the discussion of its uniqueness as
overcomplicated. Starting from $M=5$ this enabled us to change the
strategy and to continue the calculations by merely confirming the
validity of the following  general odd-dimensional formula
 \ben
 A^{(EEP)}=M(M+1),\ \ \ \  B^{(EEP)}=M(M+1)-1\cdot 2=M(M+1)- 2,\ \ \
 \een
 \be
   C^{(EEP)}=M(M+1)-2\cdot 3,\ \ \ \
 D^{(EEP)}=M(M+1)-3\cdot 4, \  \ldots\,.
 \label{sulika}
 \ee
This formula is our second main result.

In order to complete the parallels with the previous section, let
us finally recollect the universal elipsoidal-surface embedding
(\ref{star}) of the domains  ${\cal D}^{(2K)}$ and emphasize that
its present odd-dimension analogue is even simpler. Indeed,
returning once more to all the $M\leq 4$ calculations of this
section we arrive at the extrapolation formula
 \be
 A+B+C+D+\ldots +Z
  =\frac{2\,M^3+3\,M^2+M}{3}\,
 \label{stardust}
 \ee
the validity of which is very easily confirmed (and was confirmed)
at a number of higher integers $M>5$. Its alternative arrangement
reads
 \be
 a^2+b^2+\ldots +z^2 \leq
 \frac{2\,M^3+3\,M^2+M}{3}\,
 \label{stardustbin}
 \ee
showing that every quasi-Hermiticity domain ${\cal D}^{(2M+1)}$ is
circumscribed by a certain minimal hypersphere, with the mutual
intersections lying precisely at the $2^M$ EEP points with the
coordinates $ a^{(EEP)} =\pm \sqrt{M(M+1)}$, $\ b^{(EEP)}=\pm
\sqrt{M(M+1)-2}$ etc.

\section{Summary}

We introduced a class of the tridiagonal and up-down symmetrized
matrix chain models $H^{(N)}$, the spectrum of which remains
equidistant in the decoupled limit. We believe that beyond their
above-mentioned direct connection to physics of harmonic
oscillators exposed to a small finite-dimensional perturbation,
another interesting source of their possible future physical
applicability could be sought in the equidistance of spectra of
certain {\em manifestly finite-dimensional} spin-chain models
possessing equidistant spectra (cf., for illustration, the
Polychronakos' SU(N) model~\cite{Polychronakos} or its
supersymmetric SU(m$|$n) generalization~\cite{Basu} etc).

At any dimension $N$ of our Hamiltonians  $H^{(N)}$ we determined
the coordinates of all the EEP (= extreme exceptional point)
$N-$plets of the matrix elements $a^{(EEP)}, b^{(EEP)}, \ldots,
z^{(EEP)}$, the choice of which leads to the maximal, $N-$fold
degeneracy of the $N-$plet of the real energy levels pertaining to
the underlying model.

At $N=2M+1$ the latter EEP values are ``maximal" in the sense of
the norm defined as a square root of the sum of their squares. The
same comment applies at the even dimensions $N=2K$ after a slight
modification of the norm taking just one half of the value of the
``central" coupling $a^2$ in the sum displayed in
eq.~(\ref{star}).

Some of the specific merits of our class of models may be seen

\begin{itemize}

 \item
in the ``user-friendly" tridiagonal structure of its Hamiltonians
$H^{(N)}$;

 \item
in the feasibility of an illustrative simulation of {\em all} the
possible scenarios leading to $2k-$tuple EP-like degeneracies of
the energies (followed by their ${\cal PT}-$symmetry-related
complexifications) at all the eligible multiplicities $k\leq N/2$;

 \item
in the fact that  for the latter and similar purposes the models
contain {\em precisely} a necessary {\em and} sufficient number of
free parameters;

 \item
last but not least, in an ``exact solvability"  leading to closed
formulae at all the dimensions $N$, for the EEP coordinates at
least.

\end{itemize}

\section*{Acknowledgement}

Supported by GA\v{C}R, grant Nr. 202/07/1307.



\newpage

\section*{Figure captions}

\subsection*{Figure 1. One quarter of the domain ${\cal D}^{(4)}$
(cf. paragraph \ref{ctyrak}) }

\subsection*{Figure 2. A magnified spike of the domain ${\cal D}^{(4)}$ }

 
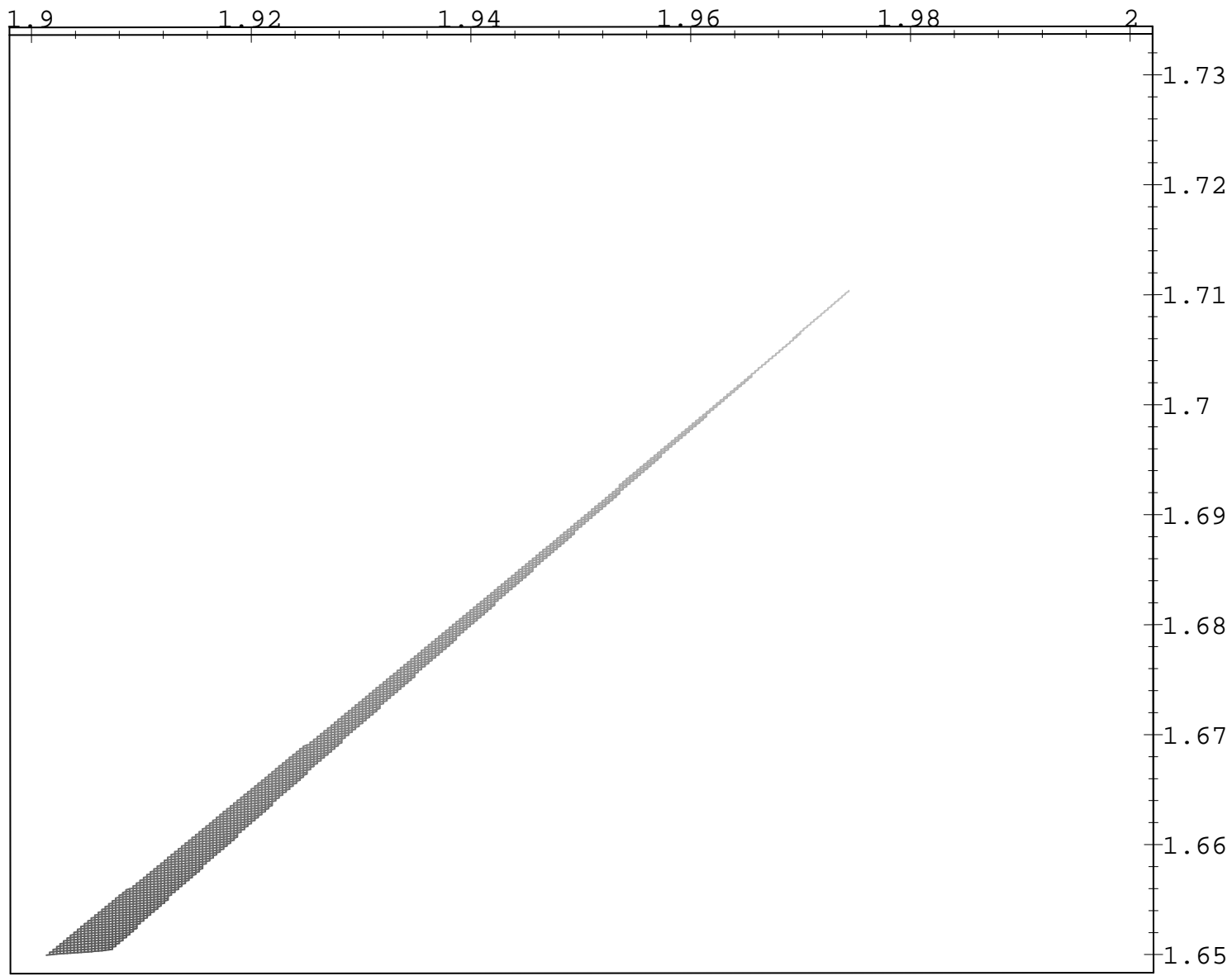
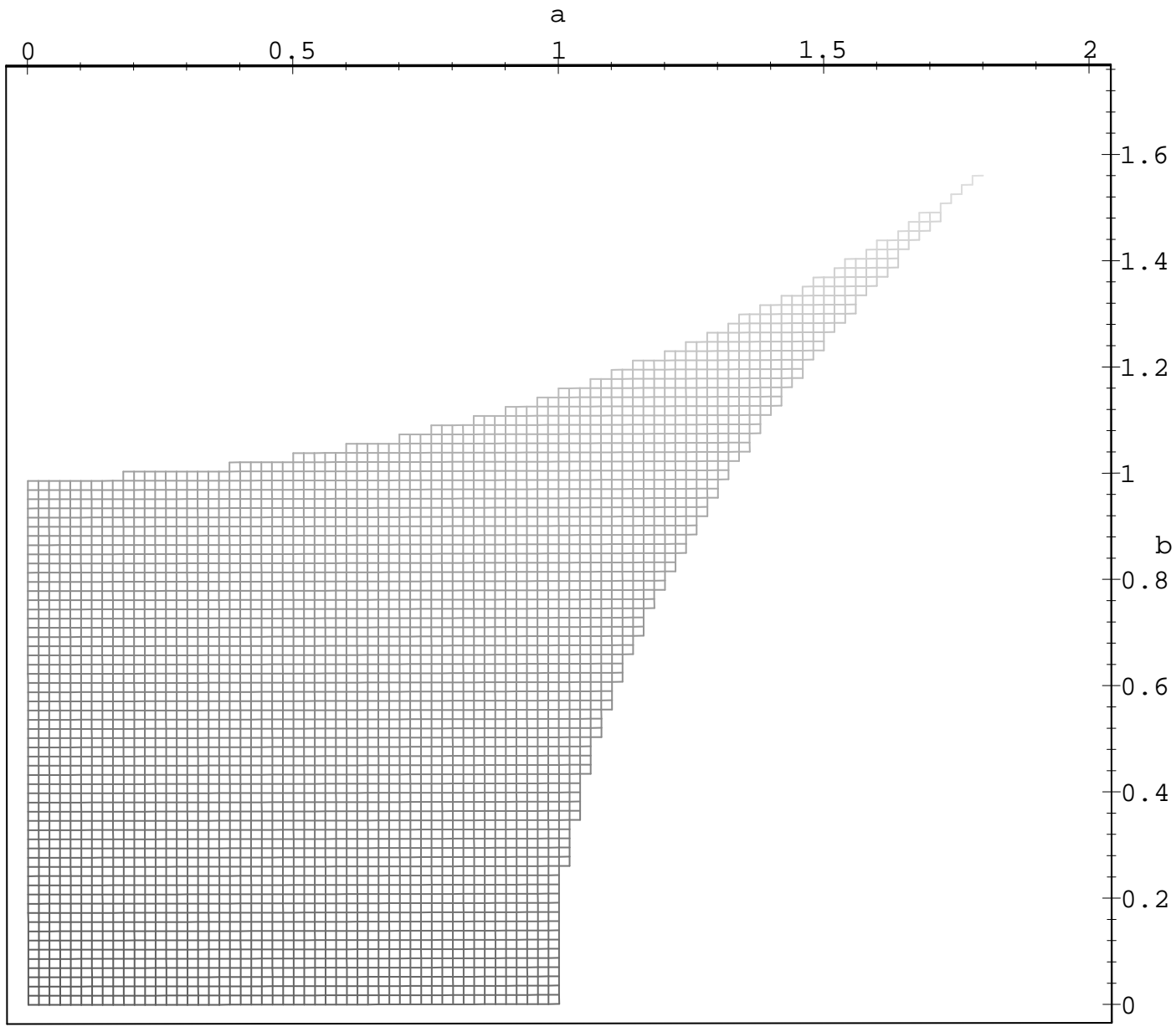
\end{document}